\documentclass[superscriptaddress,longbibliography,floatfix, nobibnotes, reprint, %endfloats
]{revtex4-2}
\usepackage{upgreek}
\usepackage{amsmath,braket}
\usepackage{graphicx}
\usepackage{mathtools}
\usepackage{color}
\usepackage{xcolor}
\usepackage{setspace}
\newcommand{\RNum}[1]{\uppercase\expandafter{\romannumeral #1\relax}}
\usepackage{comment}
\usepackage{soul}

\makeatother

\begin{document}
\title{The Mollow triplets under few-photon excitation}

\author{Bang~Wu}
\thanks{These authors contributed equally.}
\affiliation{Beijing Academy of Quantum Information Sciences, Beijing 100193, China}

\author{Xu-Jie~Wang}
\thanks{These authors contributed equally.}
\affiliation{Beijing Academy of Quantum Information Sciences, Beijing 100193, China}

\author{Li~Liu}
\affiliation{Beijing Academy of Quantum Information Sciences, Beijing 100193, China}

\author{Guoqi~Huang}
\affiliation{Beijing Academy of Quantum Information Sciences, Beijing 100193, China}
\affiliation{School of Science, Beijing University of Posts and Telecommunications, Beijing 100876, China}

\author{Wenyan~Wang}
\affiliation{Beijing Academy of Quantum Information Sciences, Beijing 100193, China}

\author{Hanqing~Liu}
\affiliation{State Key Laboratory of Superlattices and Microstructures, Institute of Semiconductors, Chinese Academy of Sciences, Beijing 100083, China}
\affiliation{Center of Materials Science and Optoelectronics Engineering, University of Chinese Academy of Sciences, Beijing 100049, China}

\author{Haiqiao~Ni}
\affiliation{State Key Laboratory of Superlattices and Microstructures, Institute of Semiconductors, Chinese Academy of Sciences, Beijing 100083, China}
\affiliation{Center of Materials Science and Optoelectronics Engineering, University of Chinese Academy of Sciences, Beijing 100049, China}

\author{Zhichuan~Niu}
\affiliation{State Key Laboratory of Superlattices and Microstructures, Institute of Semiconductors, Chinese Academy of Sciences, Beijing 100083, China}
\affiliation{Center of Materials Science and Optoelectronics Engineering, University of Chinese Academy of Sciences, Beijing 100049, China}

\author{Zhiliang~Yuan}
\email{yuanzl@baqis.ac.cn}
\affiliation{Beijing Academy of Quantum Information Sciences, Beijing 100193, China}

\date{\today}

\begin{abstract}
{Resonant excitation 
is an essential tool in the development of semiconductor quantum dots (QDs) for quantum information processing.
One central challenge is to enable a transparent access to the QD signal without post-selection information loss.
A viable path is through cavity enhancement, which has successfully lifted the resonantly scattered field strength over the laser background under \emph{weak} excitation. Here, we extend this success to the \emph{saturation} regime using a QD-micropillar device with a Purcell factor of 10.9 and an ultra-low background cavity reflectivity of just 0.0089. 
We achieve a signal to background ratio of 50 and an overall system responsivity of 3~\%, i.e., we detect on average 0.03 resonantly scattered single photons for every incident laser photon.    
Raising the excitation to the few-photon level, the QD response is brought into saturation where we observe the Mollow triplets as well as the associated cascade single photon emissions, without resort to any laser background rejection technique. Our work offers a new perspective toward QD cavity interface that is not restricted by the laser background.}
\end{abstract}

\maketitle

\normalsize

Semiconductor quantum dots (QDs) are an attractive platform for engineering high performance quantum emitters~\cite{tomm2021bright,wang2019} owing to their solid-state nature and prospect for monolithic integration into a larger system with tailored optical properties~\cite{Prtljaga2016,Tiranov2023}.
Under resonant excitation, they have been demonstrated to exhibit a plethora of quantum electrodynamic phenomena~\cite{muller2007resonance,press2008complete,flagg_resonantly_2009,liu2018,loo2012,Wei2021} that could underpin a wide range of quantum information applications. 
However, the quantum light by resonant scattering is often accessible just in the degree of freedom that is orthogonal to the driving field, e.g.,  in a cross-polarization configuration\cite{tomm2021bright,wang2019,liu2018,Vamivakas2009,wei2014temperature,Unsleber2015}. The incomplete access not only halves the collection of the resonantly scattered signal but also, more detrimentally, precludes  QD's applications for polarization-based protocols such as generation of photonic cluster states using charged excitons~\cite{Lindner_2009}.
Solution to this problem lies within cavity enhancement of light-matter interaction, which can reduce the laser excitation power by orders of magnitude and has thus far enabled a photon filter device~\cite{Bennett2016,deSantis2017} that transmits or reflects a weak coherent source according to its photon number~\cite{chang2007}. 
Here we extend this success towards \emph{few-photon} excitation regime in a QD-micropillar system with a record low cavity reflectivity and demonstrate the first observation of the Mollow triplets for QDs without resort to a laser background rejection technique.  
Due to the intimate connection between the Mollow triplets and Rabi oscillation, translating into pulsed excitation could lead to generation of deterministic single photons with arbitrary polarization encoding.

\begin{figure*}[!t]
\centering
\includegraphics[width=10 cm]{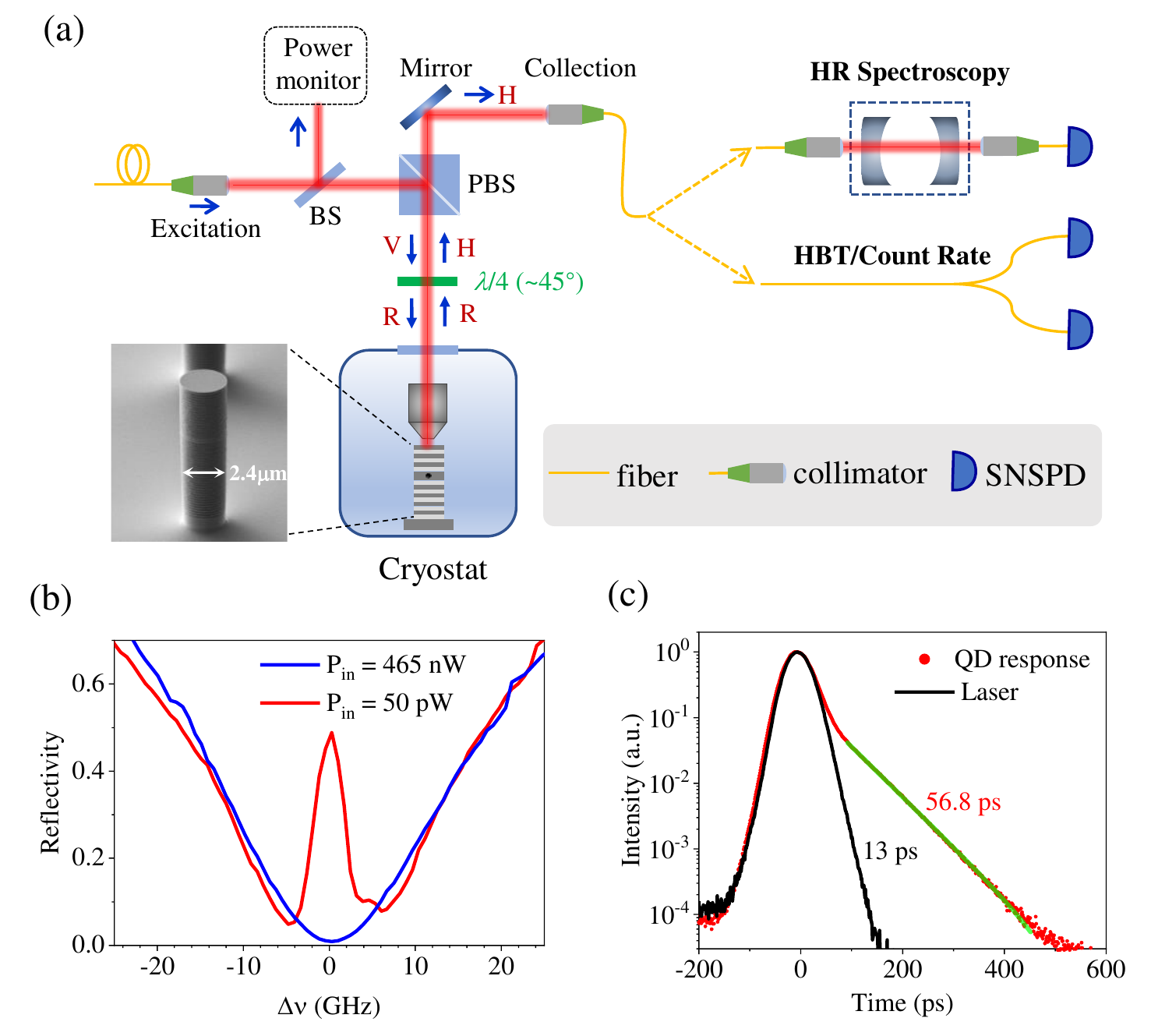}
\caption{Experimental setup and QD-cavity characterisation. (a) Setup for direct collection of resonance fluorescence without a cross-polarization filter. The combination of the polarization beam splitter (PBS) and a $\sim$45$^\circ$ $\lambda/4$-plate forms a single polarization circulator. The QD signal is collected into a single mode fiber, which can feed to setups for high resolution (HR) spectroscopy, the QD signal count rate, and the second-order correlation function measurement using a Hanbury-Brown Twiss (HBT) setup. (b) Reflectivity spectra of the cavity obtained under strong (blue line) and weak (red line) continuous-wave excitation. (c) Time-resolved resonance fluorescence (red dots) of the QD device, as well as the instrument response function (black line) measured for 5~ps laser pulses. The green line is an exponential fit with a decay time of $T_1=56.8$ ps.}
\label{fig:setup}
\end{figure*}

When resonantly driven by a strong continuous-wave (CW)
optical field, the ground and excited states of a two-level emitter (TLE) become coupled by 
electric-dipole interaction, resulting its fluorescence spectrum to exhibit a distinctive three-peak structure known as the `Mollow triplets’, as theoretically derived for atoms by Mollow in 1969~\cite{mollow_power_1969}. This behaviour is best understood 
as `dressed-states’ with their new eigenstates split by the Rabi energy ($\hbar\Omega$) that is proportional to the amplitude of the driving optical field. Among four possible transitions between the dressed states, two are degenerate at the bare TLE transition ($E_0$) while the other two are displaced by the Rabi energy to $E_0 \pm \hbar\Omega$. The spectral signature consists of three lines weighing 1:2:1 in intensity. 
%%%%
Following the first experiment on atomic sodium~\cite{wu1975},  the Mollow triplets have been routinely investigated across gaseous systems~\cite{ng2022,sterk2012,ortiz2019mollow} to solid systems~\cite{astafiev2010resonance,muller2007resonance,wrigge2008efficient,flagg_resonantly_2009}.  
For semiconductor QDs, the optical excitation power required to observe the Mollow triplets is often in the microwatt~\cite{Vamivakas2009,Ulrich2011,wei2014temperature} or even sub-milliwatt~\cite{flagg_resonantly_2009,ulhaq_cascaded_2012,Unsleber2015} region, which is far beyond the single photon level.
Consequently, all these QD experiments used either orthogonal-geometry~\cite{flagg_resonantly_2009,ulhaq_cascaded_2012,Ulrich2011} or cross-polarization techniques~\cite{Vamivakas2009,wei2014temperature,Unsleber2015} to suppress the laser background.

Grown by molecular beam epitaxy, our sample contains a single layer of low density InAs QDs at the center of a GaAs $\lambda$-cavity with 18 (30.5)  GaAs/Al$_{0.9}$Ga$_{0.1}$As mirror pairs as its  upper (lower) distributed Bragg reflector (DBR).  Micropillar arrays with diameters between 2 and 3~$\mu$m were fabricated using photolithography and inductive coupled plasma etching.
The sample is kept in a close-cycled cryostat and investigated using a confocal microscopy setup as schematically shown in Fig.~\ref{fig:setup}(a). We combine a polarization beam splitter (PBS) and a $\lambda/4$ waveplate to form a single-polarization optical circulator that routes the outgoing resonantly scattered signal of the same polarization as the incident laser for collection by a single mode fiber. The $\lambda/4$ waveplate is slightly off from 45$^\circ$ to compensate the birefringence of the device and thus maximize the QD response.  Spectroscopy or the second-order intensity correlation function are measured with superconducting nanowire single photon detectors (SNSPDs).  A scanning Fabry-Perot cavity is used for high resolution fluorescence spectroscopy, and grating-based filters are chosen to select photons from Mollow triplet sidebands for correlation measurements.

We use a pillar microcavity of 2.4~$\mu$m diameter (Fig.~\ref{fig:setup}(a), inset). It contains a quantum dot whose neutral exciton (X) transition at 911.55~nm is in resonance with the HE$_{11}$ mode of the device at 10.9~K. The cavity mode has a quality factor of around 9350, as characterised by scanning
reflectance spectroscopy (Fig.~\ref{fig:setup}(b)) under strong excitation (465~nW).
Strikingly, the reflectance spectroscopy reveals a 
low reflectivity of $R_{min}$ = 0.0089 at the cavity resonance wavelength, indicating most of the excitation light transmitting through the cavity. We note that this reflectivity is about 10 times lower than reported previously for micropillar devices~\cite{Bennett2016,deSantis2017}. 
Reducing the excitation power to 50~pW, we observe a strong reflectance peak arising from the cavity-enhanced coherent scattering~\cite{Bennett2016} of the QD. 
We obtain a peak reflectivity of $R_{max}$ = 0.49, implying a record contrast of $R_{max}/R_{min} = 55$ over the cavity background. 
We note that this contrast is limited by the QD blinking. Under 50~pW excitation, we observe telegraphic type dependence of the QD signal count rate (see Supplemental Document), from which we can revise the QD reflectance to 0.66 and can potentially improve the signal contrast to 74 when using blinking-free QDs~\cite{zhai2020low,cao2022surface}.  

Figure~\ref{fig:setup}(c) shows time-resolved resonant fluorescence intensity under excitation of 80~MHz, 5~ps Ti:Sapphire laser pulses.  As the laser linewidth (0.23~nm) is considerably broader than the cavity's (0.097~nm), we observe a pronounced background at zero time delay before the exponential decay of the QD scattered signal. We extract a cavity-enhanced exciton lifetime of $T_1 = 56.8$~ps and a Purcell factor of 10.9 by comparing it to the lifetime of 674.4~ps measured with the transition tuned out of the cavity resonance via temperature. Non-resonant excitation results are shown in Supplemental Document.  
To obtain the exciton dephasing time $T_2$, we launch the CW scattered signal into a Michelson interferometer and measure the first-order interference visibility as a function of the interferometer delay.  As typical~\cite{matthiesen2012subnatural}, two decay components are observed, with the rapid (slow) component attributed to the exciton dephasing (the laser coherence).  We obtain $T_2=103.5$~ps.  
This value is comparable, but slightly longer than, the value of 79.1~ps inferred from the linewidth of the QD reflectance spectrum (Fig.~1b). 
The discrepancy is attributed to the QD fine-structure splitting of $\sim$1~GHz which causes an underestimation of the coherence time.  
The value of $T_2/2T_1 = 0.91$ is close to an ideal TLE, suggesting an insignificant pure dephasing. 

\begin{figure}[htbp]
\centering

\includegraphics[width=8 cm]{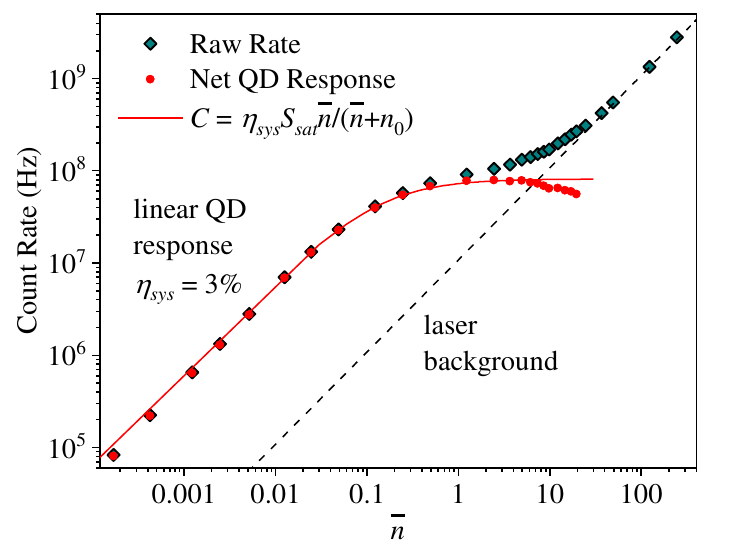}
\caption{ Count rate dependence. The QD response count rate is plotted as a function of the average number of photons per exciton lifetime ($T_1$). Here, $\bar{n}=1$ corresponds to an excitation power of 3.84~nW. At high count rate regime, we attenuate the photon signal before entering the SNSPDs to avoid detector saturation.}
\label{fig:2}
\end{figure}

Figure~\ref{fig:2} shows the power dependence of the scattered count rate ($C$) under resonant excitation by a CW laser of 100~kHz linewidth.  For clarity, we use average flux per exciton lifetime $\bar{n} = P_{in}T_1/h\nu$ to depict the excitation power, where $P_{in}$ is the optical power incident upon the sample surface,
$h$ is the Planck constant and $\nu$ the laser optical frequency.  
As compared with intra-cavity number usually used in literature~\cite{loo2012}, the parameter $\bar{n}$ can be rigorously calibrated and allows a conservative quantification of light-exciton coupling. The measured $C\sim\bar{n}$ relationship can be summarised into two linear sections  sandwiching a quasi-plateau between $\bar{n}\in [0.1, 10]$.
In the first linear region, we observe a strong anti-bunching in the second-order correlation function measurement using the HBT setup and obtain $g^{(2)}(0) = 0.03$ at $\bar{n}=0.02$, proving the quantum nature of the scattered field as a TLE can absorb and re-emit only one photon at a time~\cite{chang2007}.  
The absolute responsivity, defined as the ratio of the photon count rate to the incident photon rate ($P_{in}/h\nu$), is characterised to be $\eta_{sys} = 3$ \% for the linear QD response.  This responsivity can potentially increase to 14.4~\%, if we correct for losses by the optics and the QD blinking (see Supplemental Document for details).

As shown in Fig.~\ref{fig:2},  the excellent linearity of the laser background (dashed line) allows a faithful subtraction off the raw signal and exposes more clearly the saturation behavior of the net QD response (red dots).   The count rate 
dependence can be well simulated using the equation~~\cite{ng2022}
\begin{equation}
C = \eta_{sys}S_{sat}\frac{\bar{n}}
{(\bar{n}+n_0)},
\end{equation}
which contains two free fitting parameters:$S_{sat}=2.716\pm0.039$~GHz the saturated photon rate scattered out of the cavity and $n_{0} = 0.125 \pm 0.007$ the saturation incident photon flux.   We note that the obtained $S_{sat}$ is a factor of 3.2 ($\sim\pi$) %4 pm 0.04$ 
lower than one would expect from the conventional equation $S_{sat} = 1/2T_1$~\cite{ng2022}.  
We have ruled out a calibration error in the incident flux, but the exact cause for the discrepancy remains a puzzle. 
We obtain a saturation count rate ($\eta_{sys}S_{sat}$) of $81.49$~MHz at an incident flux of  $\bar{n} \simeq 1.0$, and this saturation extends to a high flux of $\bar{n} \simeq 4$ before falling gently as the flux further increases.  This fall could be due possibly to the interference between the QD response and the reflected laser background, which would become most pronounced when the two interfering components are comparable in intensity.  

\begin{figure*}[!t]
\centering

\includegraphics[width=10cm]{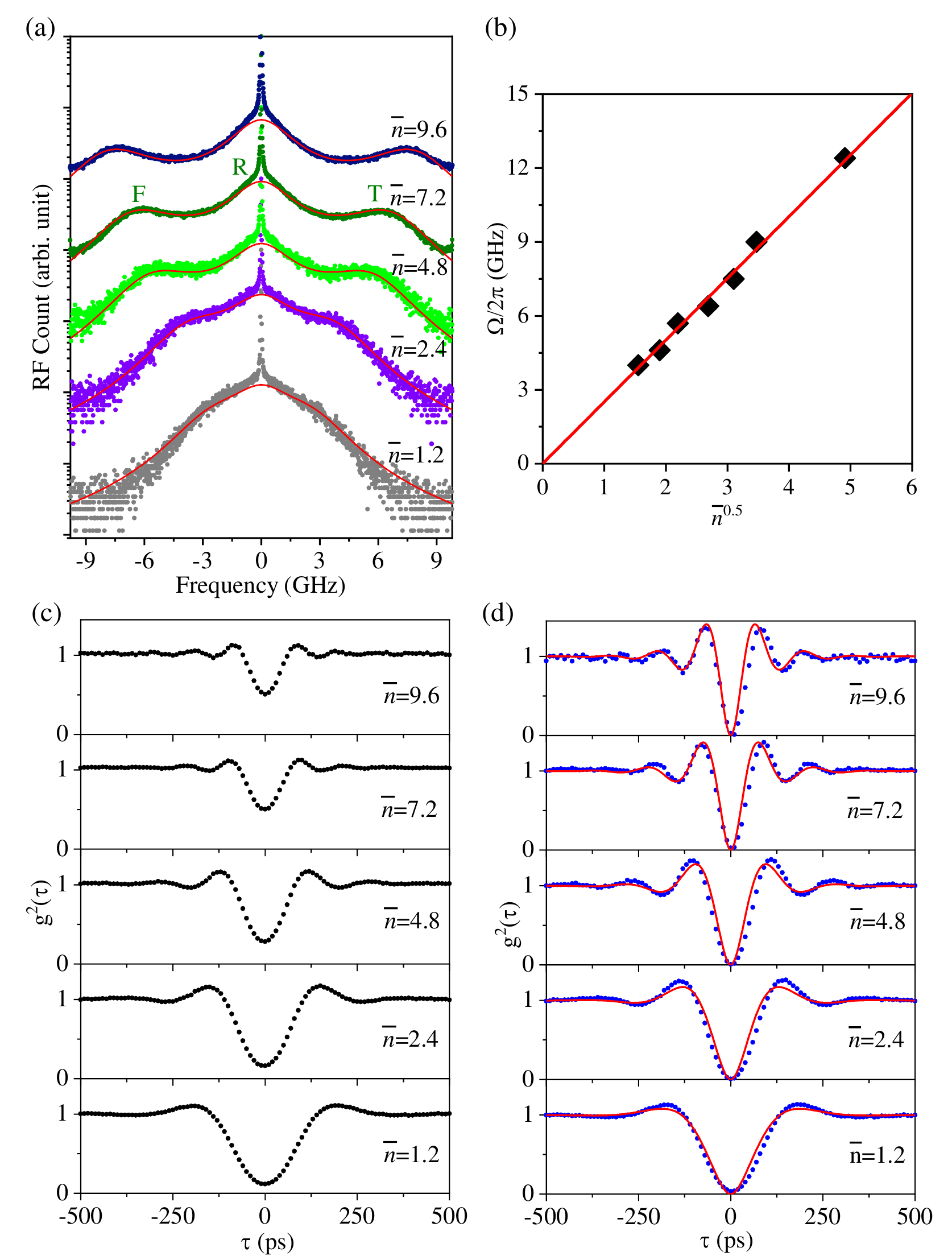}
\caption{The Mollow triplets. (a) Spectra of the Mollow triplets for different incident photon flux.  For clarity, the data are normalized and offset in log-scale. Lines are simulations using Eq.~1. 
(b) Rabi frequencies (dots) extracted from a together with a linear fit (red line). (c) The second-order correlation measurements.  (d) The Second-order correlation data after background subtraction and deconvolution of the IRF, together with simulation using Eq.~2.
In the simulations, we use $T_1$=56.8~ps and $T_2$ = 103.5~ps.}
\label{fig:3}
\end{figure*}

In the count rate plateau region, the QD signal remains dominant over the laser background, thus allowing the first investigation of the Mollow triplets for QDs without using a background rejection technique.  
As shown in Fig.~\ref{fig:3}\textbf{c}, the QD device maintains its quantum nature as evidenced by the measurements of the second-order-correlation function for various incident photon flux.  At the lowest flux of $\bar{n} = 1.2$,  $g^{(2)}(\tau)$ exhibits mostly a simple dip with a small $g^{(2)}(0)$ of 0.11, which is a clear indication of photon anti-bunching phenomenon.  By increasing the incident flux to a few photons, oscillations in the $g^{(2)}(0)$ value emerge and have a shortening oscillation period, which we attribute to the QD undergoing Rabi oscillations within its exciton lifetime.  At $\bar{n} = 7.2$, the $g^{(2)}(0)$ value still overcomes  the classical limit (0.5).

Correspondingly in the frequency domain, a series of Mollow-triplets spectra for different excitation powers are clearly observable, see Fig.~\ref{fig:3}(a). Here, a scanning Fabry-Perot cavity with a finesse $\sim1000$ and a linewidth 15 MHz is used. The sharp peak at the 0-offset arises from the bare cavity reflection, superimposing on the broad features of the QD response. 
As the incident flux increases, the broad feature evolves from a single peak to shoulders and finally sidebands with an increasing separation. The three broad peaks are identified as the `fluorescence’ line (F), `Rayleigh’ line (R), and `three photon’ line (T) respectively.  For $\bar{n} \geq 2.4$, we can extract the corresponding  Rabi frequencies from the splittings between the sidebands, and confirm a linear relationship of $\Omega \propto \bar{n}^{1/2}$ (Fig.~\ref{fig:3}(b)), as expected for the Mollow triplets.  
The Mollow triplets power spectra can be reproduced using the equation~\cite{flagg_resonantly_2009}:

\begin{widetext}
\begin{flalign}  S(\Delta\omega)=\frac{n_\infty}{\pi}[{\frac{1}{2}\frac{1/T_2}{(\Delta\omega)^2+1/T_2^2}}
+\frac{n_\infty}{\Omega^2}(\frac{A\eta/2-B(\Delta\omega-\mu)/8\mu}{(\Delta\omega-\mu)^2+\eta^2}+\frac{A\eta/2+B(\Delta\omega+\mu)/8\mu}{(\Delta\omega+\mu)^2+\eta^2})],
 \label{eq:2}
 \end{flalign}
 \end{widetext}
 
\noindent where $\Delta\omega/2\pi$ is the frequency detuning from the central QD exciton resonance,  $n_\infty=(\Omega^2T_1/T_2)/[2(\Delta\omega^2+T_2^{-2}+\Omega^2T_1/T_2)]$ the steady-state population,
and parameters A, B, $\eta$ and $\mu$ are functions of $T_1$ and $T_2$: $A=\Omega^2+(1/T_1-1/T_2)/T_1$,
$B=2\Omega^2(3/T_1-1/T_2)-2(1/T_1-1/T_2)^2/T_1$, $\eta=(1/T_1+1/T_2)/2$ and $\mu=\sqrt{\Omega^2-(1/T_1-1/T_2)^2/4}$.  
Using experimentally measured $T_1$ and $T_2$ as well as the extracted Rabi frequencies (Fig.~\ref{fig:3}(b)), we can faithfully reproduce the Mollow spectra in simulations as shown by in Fig.\ref{fig:3}(a) (red lines).

For resonant scattering of an ideal TLE system, the $g^{(2)}$ function can be simulated using~\cite{flagg_resonantly_2009}
\begin{equation}
g^{(2)}(\tau)=1-\exp(-\eta\lvert{\tau}\rvert)[\cos(\mu\lvert{\tau}\rvert)+\eta/\mu \sin(\mu\lvert{\tau}\rvert)].
\label{eq:3}
\end{equation}
To allow a direct comparison, we remove the laser background from the raw experimental data and de-convolute the HBT instrument response function \iffalse, which is described in the Supplemental Document\fi.  As shown in Fig.~\ref{fig:3}(d), the processed data (dots) and their simulation (red lines) are in excellent agreement.

\begin{figure*}[htbp]
\centering

\includegraphics[width=14cm]{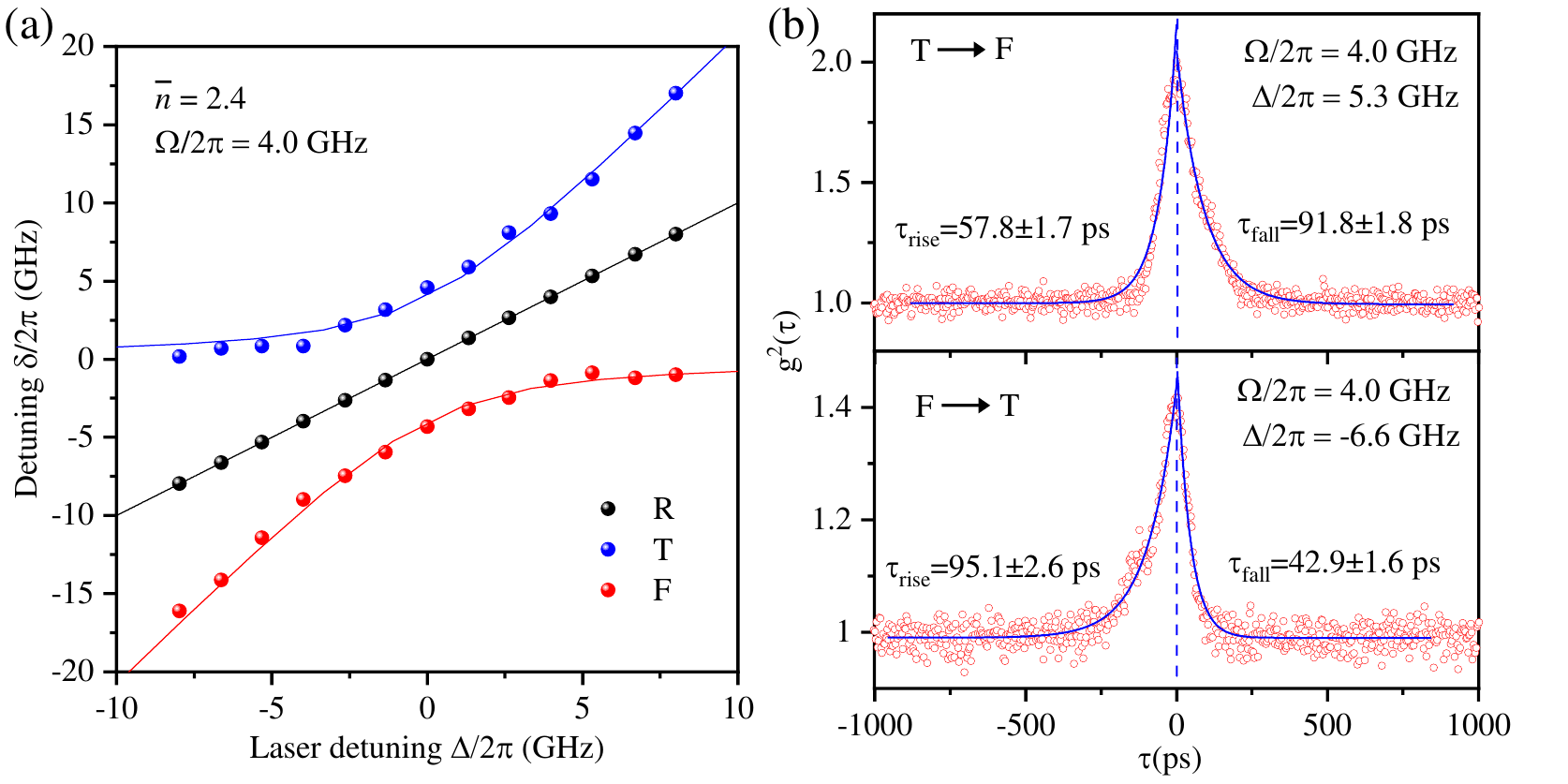}
\caption{\label{fig:4} Cascaded single-photon emissions from two opposite Mollow sidebands. (a) Peak positions of the F, R, and T components as a function of the laser detuning.  Solid lines are fits using  $\delta=\Delta\pm\sqrt{\Omega^2+\Delta^2}$.
(b) Normalized photon correlation between the T photon (‘start’) and F photon (‘stop’), for the laser is blue (red)-detuned by $\Delta/2\pi=5.3$ ($-6.6$) GHz from the exact QD resonance. 
}
\end{figure*}

The Mollow triplets are also investigated as a function of the laser frequency detuning. The result is plotted in Fig.~\ref{fig:4}(a). All data were measured at the same incident flux $\bar{n} = 2.4$,  
corresponding to a bare Rabi frequency of 4~GHz.  The R peak frequency shifts with the laser frequency, while the F and T transitions remain  symmetric with respect to the R transition. The generalized Rabi splitting is $\Omega_g=\sqrt{\Omega^2+\Delta^2}$, where $\Delta/2\pi$ is the laser detuning frequency.
For large detunings, the symmetry of the dressed states population is significantly broken, as the emission process of the R and T photons now have a preferred order~\cite{ulhaq_cascaded_2012,nienhuis_spectral_1993}.
For $\Delta\gg0$, the time-ordered cascaded
emission of a T photon ‘heralding’ the F transition results in an
asymmetric photon-bunching signature of the Mollow sidebands in the
cross-correlation experiment, as demonstrated in Fig.~\ref{fig:4}(b). Here, we use two grating-based filters (8~GHz bandwidth) to select the F or T photons from the incoming scattered signal before directing to separate single photon detectors. 
The normalized correlation is fitted by two exponentials, with time constants of $\tau_{rise}= 57.8$ ps and
$\tau_{fall}= 91.8$ ps, respectively.
In contrast, the case of strongly red-detuned
excitation ($\Delta/2\pi=-6.6$ GHz) is shown in Fig.~4b. Here, the emission order is revsed to a T photon following an F photon, resulting a rapid rise ($\tau_{rise}= 42.9$~ps) and a slower fall ($\tau_{fall}= 95.1$~ps).

In summary, we have observed the Mollow triplets at few-photon excitation as well as cascade single photon emissions from opposite side-bands in a QD-micropillar without resort to a cross-polarization setup nor other background rejection techniques. 
We attribute this success to the device's intrinsically low cavity reflectivity as well as efficient light coupling into the QD via cavity enhancement.  
A complete access to the QD signal could allow excitation and collection of photons in arbitrary polarizations, which is crucial step towards generation of photonic cluster states. 
Furthermore, observation of the Mollow triplets implies the feasibility of using deterministic pulsed excitation for efficient generation of single photons, with lossless encoding via modulating  the polarization of the excitation beam. 

%\begin{widetext}

\onecolumngrid
\section*{Supplementary}

\subsection{Non-resonant characterization}

\begin{figure}[htbp]
\centering
\includegraphics[width=12 cm]{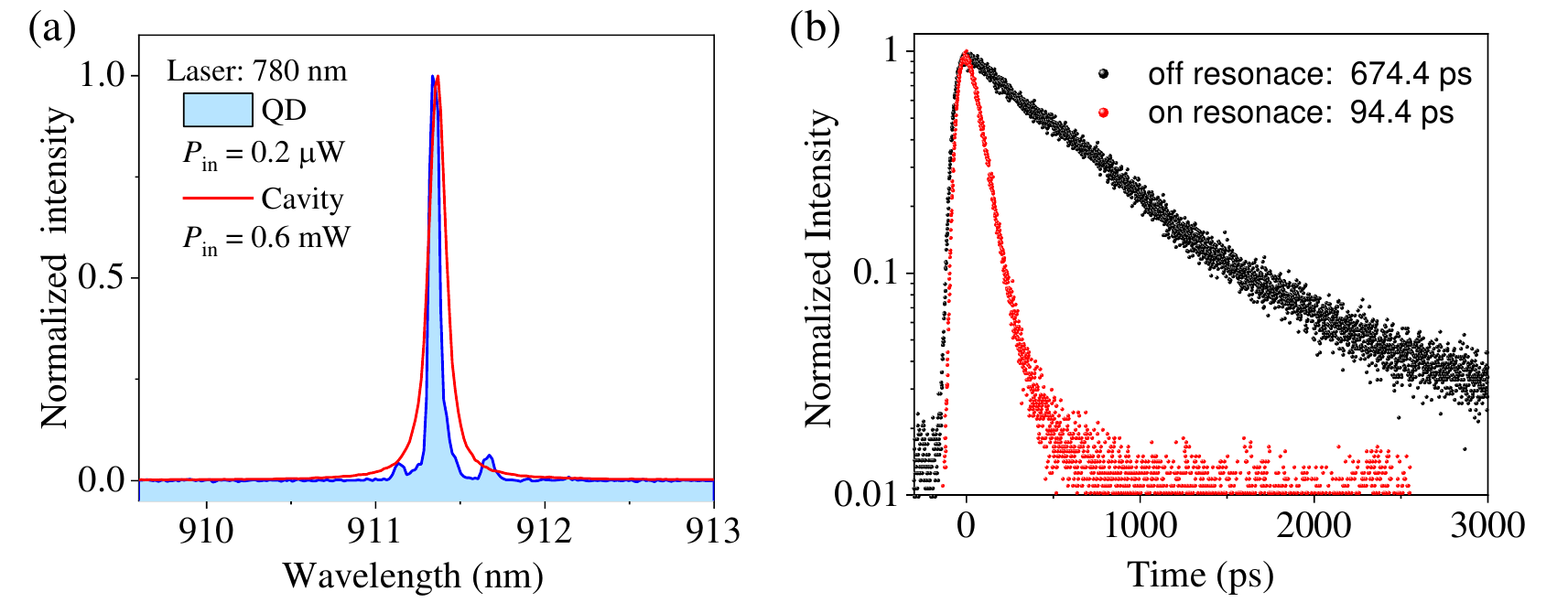}
\caption{Non-resonant excitation measurements. (a) Normalized PL spectra of the QD-micropillar QD device under low power ($0.2\mu W$) and high power (0.6~mW) excitations; (b) Time-resolved PL spectra for the QD in (black) and out of (red) the resonance with the cavity mode.
}
\label{fig:s1}
\end{figure}

Figure~\ref{fig:s1}(a) shows photoluminescence (PL) spectra of the QD-micropillar device under 780-nm CW laser excitation using a spectrometer with a resolution of 0.008~nm.
Under high power excitation (600~$\mu$W), the cavity mode dominates the spectrum  and exhibits a linewidth of 0.11~nm, corresponding to a Q factor of 8300. This Q value is slightly smaller than 
the scanning reflectance result shown in Fig.~1(b), Main Text, because of the limiting resolution of the spectrometer.  
Under a 3-orders of magnitude lower power excitation,  we observe spectral lines from individual QDs.  The QD emission line can be temperature tuned across the cavity resonance.  At 10.9~K, we observe a cavity enhanced QD PL peak featuring a spectral width of 0.06~nm.

We measure the QD exciton lifetime using 780~nm, 5~ps Ti:S laser pulses.  As shown in Fig.~\ref{fig:s1}(b), we obtain two different lifetimes:  674.4~ps when the QD is far detuned and 94.4~ps when the QD is in resonance with the cavity.  
The on-resonance lifetime is noticeably longer than the resonant excitation result of 56.8~ps obtained in Fig.~1c.  This discrepancy arises from the fact that photo-carriers under non-resonant excitation have to undergo a long relaxation process before emission~\cite{liu2018}. 
Hence, we use the resonant excitation lifetime of 56.8~ps to calculate the device Purcell factor and obtain a value of 10.9.

\subsection{Extraction of the exciton dephasing time} 

\begin{figure}[htbp]
\centering
\includegraphics[width=10cm]{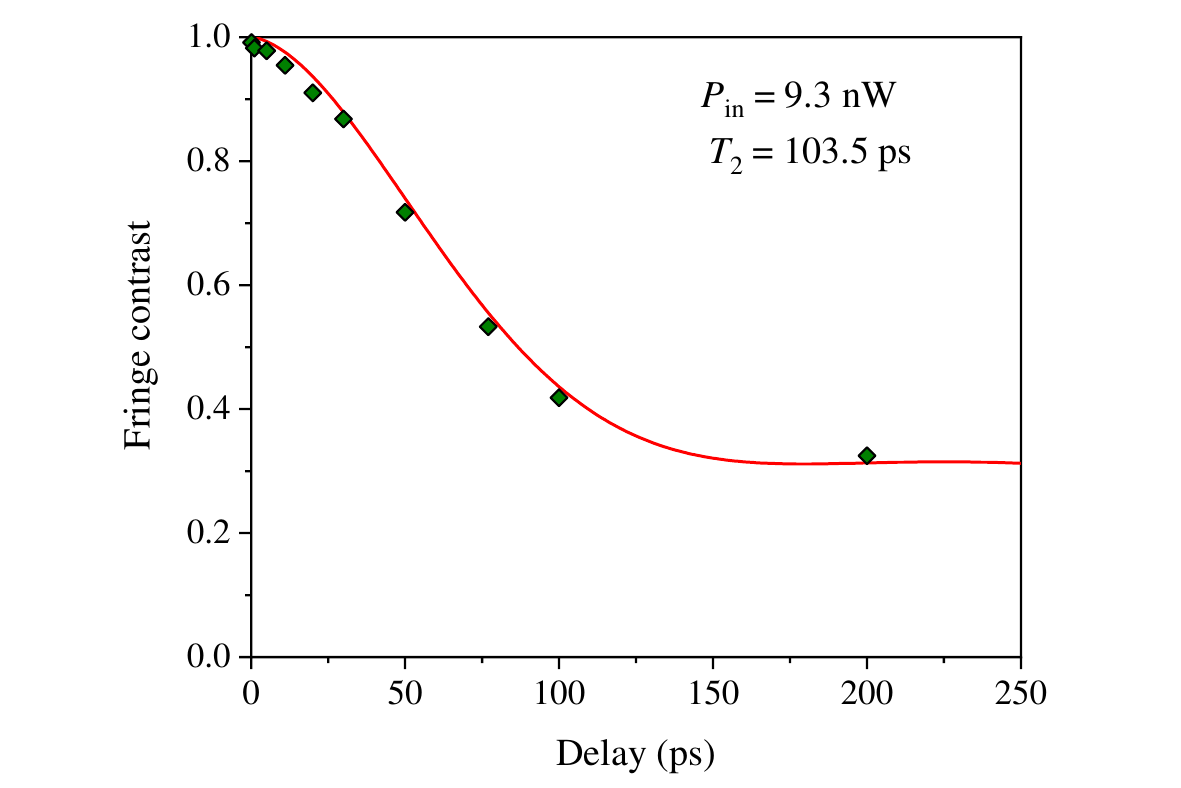}
\caption{The first-order interference visibility as a function of the Michelson interferometer differential delay. }
\label{fig:s2}
\end{figure}

We use a Michaelson interferometer with a variable arm delay to measure the exciton dephasing time $T_2$ \cite{santori2002indistinguishable}, which is extracted from the dependence of the first-order interference visibility on the arm differential delay. We choose an excitation power of 9.3~nW, which corresponds to $\bar{n}=2.4$ and is sufficiently large to discern the fast coherence decay component.  As typical~~\cite{muller2007resonance,matthiesen2012subnatural}, the rapid  (slow) decay component is attributed to the exciton dephasing (the laser coherence).  
The rapid component is expected to follow~\cite{muller2007resonance,matthiesen2012subnatural}
\begin{flalign}
g^{(1)}_{incoh}\propto&\frac{\Omega^2}{2(\Omega^2+\Gamma_1\Gamma_2)}[\frac{1}{2}e^{-\Gamma_2\tau}+e^{-1/2(\Gamma_1+\Gamma_2)\tau}(\frac{1}{2}\frac{\Omega^2+\Gamma_1\Gamma_2-\Gamma_1^2}{\Omega^2+\Gamma_1\Gamma_2}\cos{(\mu\tau)}&\nonumber\\&
-\frac{1}{4\mu}\frac{\Omega^2(\Gamma_2-3\Gamma_1)+\Gamma_1(\Gamma_2-\Gamma_1)^2}{\Omega^2+\Gamma_1\Gamma_2}\sin{(\mu\tau)})],
\label{eq.s1}
 \end{flalign}
where $\Omega$ is the Rabi frequency,  $\mu=\sqrt{\Omega^2-(1/T_1-1/T_2)^2/4}$, $\Gamma_1=1/T_1$, $\Gamma_2=1/T_2$. Using experimentally measured $T_1$ and the extracted Rabi frequencies (Fig.~3b) as well as a fixed value for $g^{(1)}_{coh}$, we can fit the exciton dephasing time $T_2=103.5$ ps as shown in Fig.~\ref{fig:s2}. 

\subsection{Signal losses}

We have established in the Main Text an overall system responsivity of 0.03, i.e., the average number of detected QD signal for each incident photon upon the device. As this responsivity directly determines future applications of a QD device for quantum information processing, it is useful to examine all extrinsic losses and thus identity the room for improvement.  As described below, the system responsivity can be revised up to 14.4~\% after correction for the losses by optical elements and the QD blinking. 

\subsubsection {Losses in the optical path}

The transmissions of all the optical elements in the signal collection path are listed in the Supplementary Fig.~\ref{fig:s3}. 
The overall transmission is 0.28. 

\begin{figure}[htbp]
\centering
\includegraphics[width=12 cm]{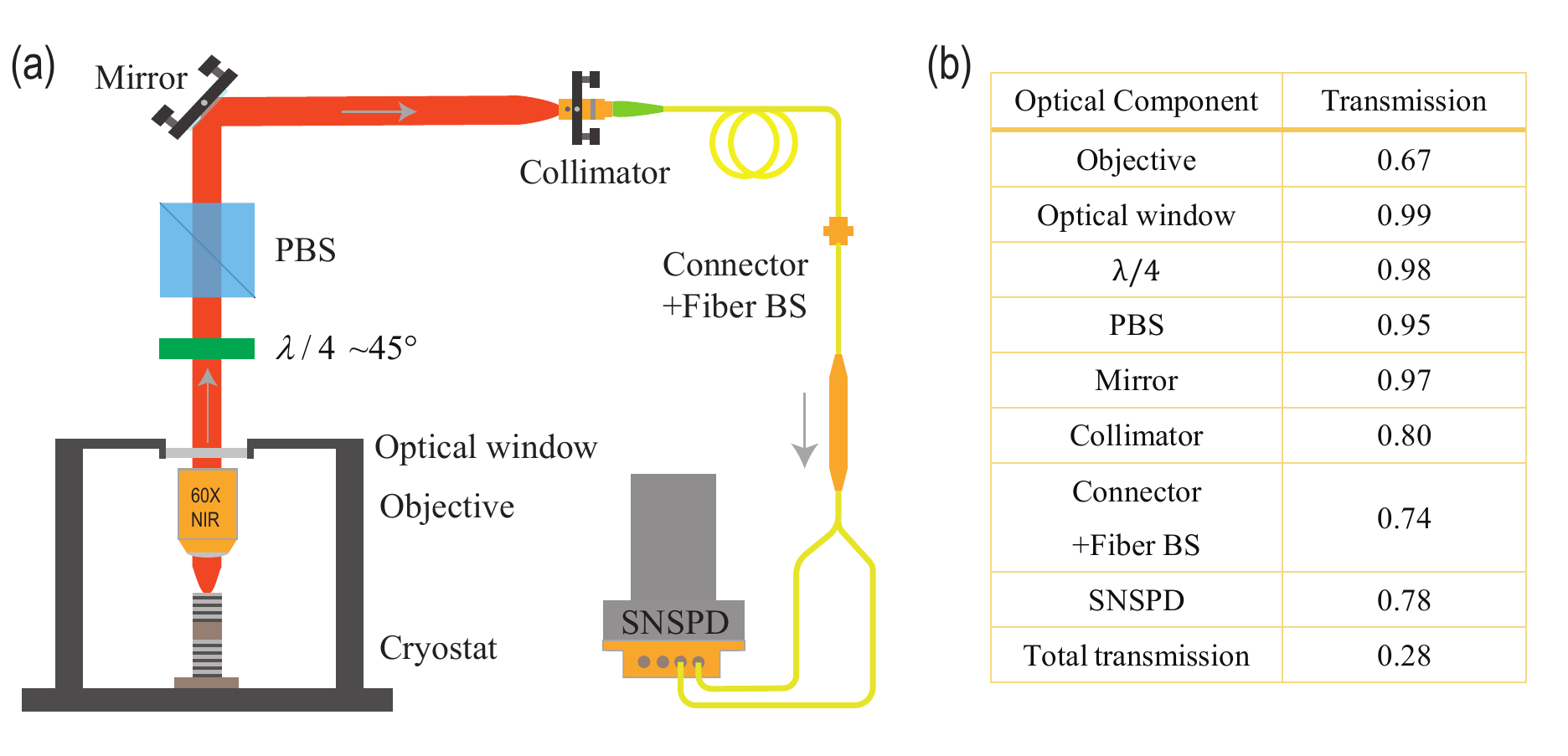}
\caption{Transmission efficiency of each component in the optical signal collection path.} 
\label{fig:s3}
\end{figure}

\subsubsection{Blinking loss}

\begin{figure}[htbp]
\centering
\includegraphics[width=13cm]{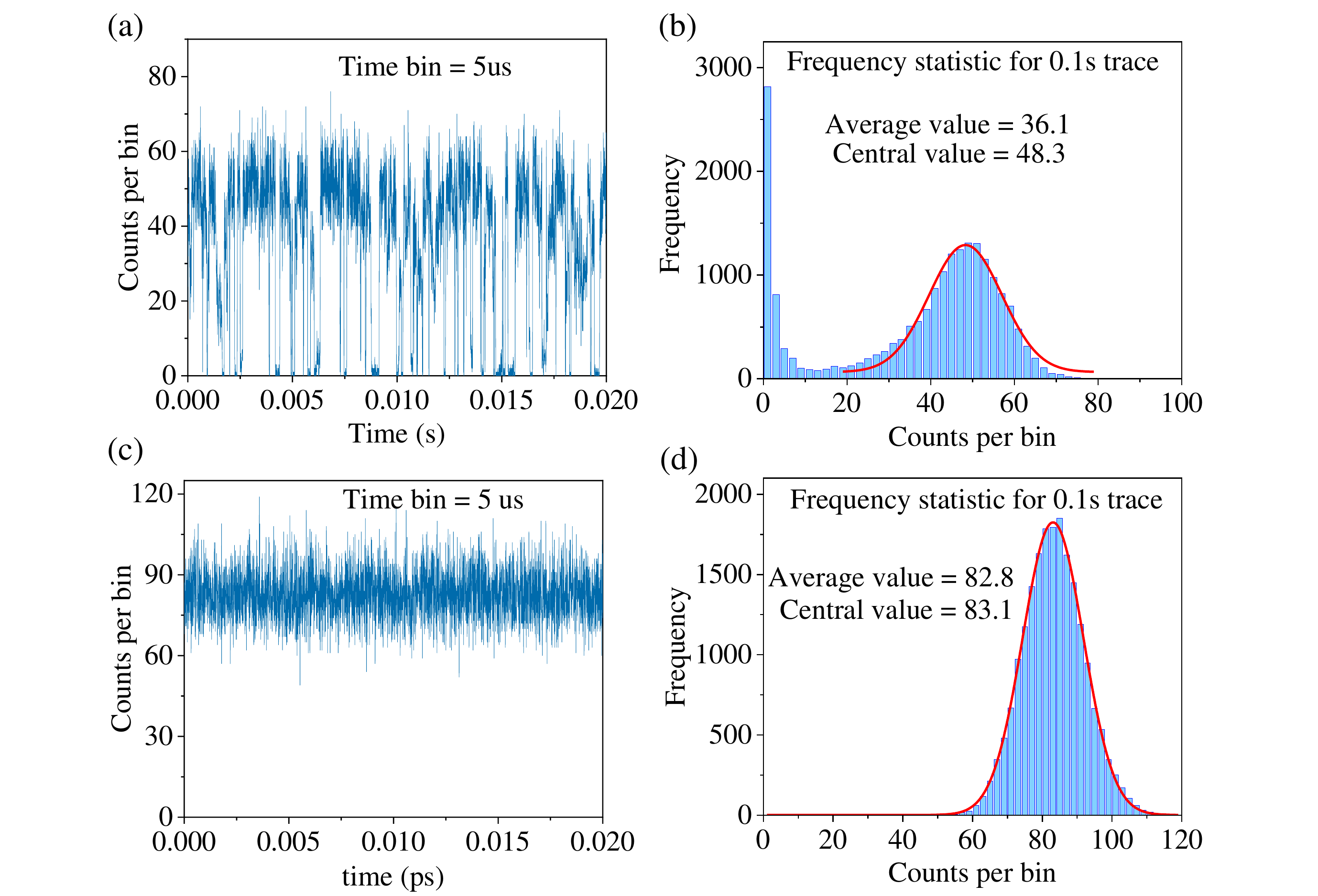}
\caption{Blinking behavior of the investigated QD. (a) the time trace of the RE countrate; (b) the corresponding frequency statistic analysis for the RE signal; (c) the time trace of the off-cavity reflected laser signal; (d) the corresponding frequency statistic analysis for the laser signal.} 
\label{fig:s4}
\end{figure}

Blinking of a QD due to its electrical environment can cause a considerable drop in the QD signal response.  As shown in Fig.~\ref{fig:s4}(a), the QD  under 50~pW CW resonant excitation blinks between `bright’ and `dark’ states over a time scale of milliseconds.  The frequency statistic analysis for a 0.1~s trace shows the `bright' (`dark') state with an average countrate of about 48.3 (0) counts/bin, as shown in Fig.~\ref{fig:s4}~(b). As a contrast, the time trace of the off-cavity reflected laser is monitored under the same condition, as shown in Fig. \ref{fig:s4}~(c,d).  The laser time trace exhibits no telegraph-type feature.  

The count ratio of the bright-state to the overall average countrate is 1.34. With this ratio, we can revise the reflectivity (Fig.~1\textbf{b}) up to 0.66 for a non-blinking QD.

%\bibliography{ref}


\clearpage

\end{document}